\documentclass{aa}

\usepackage{graphicx}

\def\approxgt{\mathrel{\hbox{\rlap{\lower.55ex \hbox {$\sim$}}
        \kern-.3em \raise.4ex \hbox{$>$}}}}
\def\approxlt{\mathrel{\hbox{\rlap{\lower.55ex \hbox {$\sim$}}
        \kern-.3em \raise.4ex \hbox{$<$}}}}

\begin{document}
\thesaurus{05(02.01.2, 11.14.1, 11.09.1:M~87, 11.03.4:Virgo, 13.25.2)}

\title{The nature of the hard X-ray power-law tail in M~87}

\author{Matteo Guainazzi\inst{1} \and Silvano Molendi\inst{2}}

\institute{
{Astrophysics Division, Space Science Department of ESA, ESTEC, Postbus 299,
NL-2200 AG Noordwijk, The Netherlands}
\and
{Istituto di Fisica Cosmica ``G.Occhialini'', Via Bassini 15, I-20133 Milano, Italy}
}
   
\offprints{M.Guainazzi}

\date{Received  ; accepted }

\maketitle

\markboth{M.~Guainazzi \& S.~Molendi}{The nature of the hard X-ray power-law tail in M~87}

\begin{abstract}

Spatially-resolved spectroscopy of the elliptical galaxy M~87 with
the MECS instrument on board BeppoSAX demonstrates that the
hard X-ray power-law tail, originally discovered by ASCA (Matsumoto et al
1996; Allen et al. 1999),
originates in the innermost 2$\arcmin$. Our results are consistent with it
being produced in an Accretion Dominated Flow, although
a substantial jet contribution cannot be ruled out. An origin
from a Seyfert-like nucleus is disfavored by our data.
As a by-product of this
result, we present an analysis of the thermal
emission coming from the center of the Virgo cluster, which exhibits a
strong  positive radial temperature gradient, along with a
radial decrease of the iron abundance.

\end{abstract}


\section{Introduction}

The first suggestion that Active Galactic
Nuclei (AGN) are fueled by accretion onto supermassive black holes
is thirty years old (Lynden-Bell 1969). The measure of
$10^6$--${\rm 10^9 M_{\odot}}$ 
black hole masses in the nuclei of
several nearby galaxies (see Ho 1998
for a review) has raised
the question of why most of them are not active. A possible solution
is provided by the so called Advection Dominated Accretion Flows
(ADAF; Rees 1982;
Narayan \& Yi 1995; Fabian \& Rees 1995). In this class
of accretion solutions, the accreting gas is so tenuous that it
cannot cool efficiently, the viscous energy is
stored in the protons as thermal energy
and eventually advected onto the nuclear
compact object. The ADAF are therefore characterized by small
radiative efficiency and accretion rates
(${\rm \dot{m} \equiv \dot{M}/\dot{M_{Edd}} < 10^{-1.6}}$; Narayan \& Yi
1995; Rees et al. 1982). At low ${\rm \dot{m}}$
the hard X-ray emission is
mainly due to a bremsstrahlung emission from a population of $\sim$100~keV
electrons,
and is therefore much harder than that expected by a standard
two phase optically-thick disk/corona scenario (Haardt \& Maraschi 1993),
which is believed to be at work in X-ray bright AGN.

The recent discovery
of power-law, hard X-ray tails in the ASCA spectra of several
elliptical galaxies and LINERS, with photon indices in the range 0.6--1.5
(Allen et al. 1999, A99; Guainazzi \& Antonelli 1999)
has provided the first X-ray observational support to the ADAF
picture.
The brightest object in the A99 sample
is the elliptical galaxy M~87 (${\rm z=0.0043}$), at the center of
the Virgo cluster, which exhibits a 20$\arcsec$ long radio and
optical jet (Biretta et al. 1991).
A difference by a factor of five
in the flux detected by {\it Spacelab 2} (Hanson et al. 1990) and
Ginga (Tanako \& Koyama 1991), suggested
a variation of the hard X-ray M~87 core emission. Actually, high-resolution
images with the ROSAT/HRI (Neumann et al. 1997; Harris et al. 1997) led
to the discovery of two soft X-ray, strongly variable (up to a factor of
two) sources, coincident with the nucleus and the so called ``knot A'',
12$\arcsec$ away from the nucleus.
The ASCA data were analyzed by several authors.
Reynolds et al. (1997) and Buote et al. (1999) did not find any
firm evidence for a power-law hard X-ray tail, with upper
limits on its 2--10~keV flux of
$8 \times 10^{-12}$~erg~cm$^{-2}$~s$^{-1}$. An analysis of the
same ASCA data by Matsumoto et al. (1996)
and A99 yielded, however, a positive
detection, with a comparable flux
and a rather poorly constrained
spectral index (${\rm \Gamma = 1.4 \pm ^{0.4}_{0.5}}$; A99).
Reynolds et al. (1999) used RXTE data to set an upper
limit of $4 \times 10^{-12}$~erg~cm$^{-2}$~s$^{-1}$.
The major difficulty
in detecting the power-law component is related to the
presence of the intense thermal emission from the core of the Virgo
cluster. The fact that this thermal emission is coming from a
multi-temperature gas (A99; Buote et al. 1999),
most likely characterized by a radial temperature and abundance gradient
(Matsumoto et al. 1996; D'Acri et al. 1998) further complicates matters.

The BeppoSAX scientific payload (Boella et al. 1997a)
is well suited to investigate
this issue in better detail, because the hard X-ray
spatial resolution provided by the Medium Energy Concentrator Experiment
(MECS; Boella et al. 1997b) is sharper
than that provided by any other mission flown before {\it Chandra}.
The results of a BeppoSAX observation of M~87
are reported in this {\it Letter}.

\section{Observation and data reduction}

M~87 was observed by BeppoSAX on 1996 July 14. In
this {\it Letter},
only data from the MECS will be considered.
The MECS (still composed of three units at the moment of the BeppoSAX
observation of M~87) is an imaging scintillation proportional counter,
with sensitive bandpass between 2 and 10.5~keV. The energy resolution is
8\% at 6~keV, and varies as ${\rm E^{-0.5}}$. 
The 80\% of the energy power is enclosed within
2$\arcmin$.5--2$\arcmin$.75.
The PSF does not significantly depend on the azimuthal angle,
as - for instance - that of ASCA.
The Low Concentrator
Spectrometer was switched off, whereas the High Pressure Gas Scintillation
Proportional Counter did not detect any emission from M~87. In the
Phoswitch Detector System (PDS; Frontera et al. 1997) a positive detection
was registered,
with a net background-subtracted count rate of $0.28 \pm 0.06$~s$^{-1}$.
However, the Seyfert~2 galaxy NGC~4388 is located about 75$\arcmin$ from the
M~87 core. A 1998 PDS observation of NGC~4388 measured a count rate of
$\simeq$2~s$^{-1}$ (M.Cappi, private communication).
Given the triangular response of the PDS collimator
(1.3$^{\circ}$ Half Width at Zero Intensity)
the PDS detection of M~87 is likely
to be substantially contaminated by NGC~4388. We will therefore not
consider these data any further. 

Data were reduced according to standard criteria, as in Guainazzi
et al. (1999). The total exposure time after data screening was 24.9~ks.
Background spectra were extracted from blank sky fields,
accumulated at the BeppoSAX Science Data Center during the first three
years of the operative life of BeppoSAX. The background subtraction does not
strongly affect the results presented in this {\it Letter}, accounting for
at most 10\% of the 9~keV count rate,
and for a rapidly decreasing 
fraction at lower energies. The total background subtracted count rate
in the innermost 4$\arcmin$
is $0.964 \pm 0.008$~s$^{-1}$.

In this {\it Letter}: uncertainties and are given at the
90\% confidence level for one interesting parameter ($\Delta \chi^2
= 2.71$); ${\rm H_0 = 50}$~Mpc~s$^{-1}$~km$^{-1}$
is assumed; energies are quoted in the source rest frame,
unless otherwise specified; errors on the best-fit values
take into account a
residual 0.8\% systematics in the gain calibration.
At a distance of M~87, 1$\arcmin$ corresponds to
7~kpc.

\section{Spatially-resolved spectroscopy}

\begin{figure}
\includegraphics*[width=8.0cm,height=8.0cm]{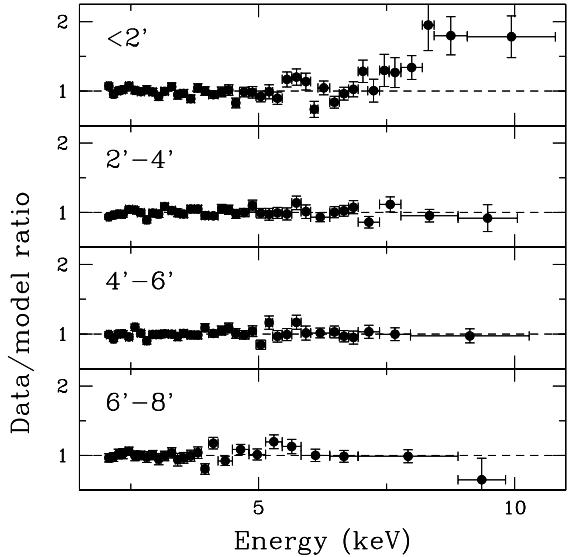}
\caption{Data/model ratio when a two temperature model
is applied to the spectra extracted in annuli with bounding radii:
0$\arcmin$-2$\arcmin$, 2$\arcmin$-4$\arcmin$, 4$\arcmin$-6$\arcmin$
and 6$\arcmin$-8$\arcmin$, from top to bottom respectively.
Each data point
corresponds to a signal-to-noise ratio $>$3}
\label{fig3}
\end{figure}

We extracted
spectra from the innermost 2$\arcmin$ around the M~87
core, and in annuli with inner--outer radii of 2$\arcmin$--4$\arcmin$,
4$\arcmin$-6$\arcmin$ and
6$\arcmin$--8$\arcmin$.
Response matrices appropriate for each spectrum were created with
the program {\sc Effarea} included in the {\sc Saxdas} data analysis
package (Molendi et al. 1999).
Several authors have pointed out the importance of fully accounting
for the complex temperature structure of the thermal emission
and for possible variations in individual elemental abundances ratios
(A99; Buote et al. 1999), especially to investigate
the presence of non-thermal hard X-ray tails.
Following their approach
(which allows us also a direct comparison with ASCA),
we fitted each spectrum with a model constituted by two
optically thin plasma emissions (the {\tt mekal} implementation in
{\sc Xspec} is used throughout this {\it Letter}). The cooler component,
which emerges below $\simeq$2~keV in ASCA, is not required by
our fits. We have therefore constrained its temperature,
${\rm kT_{soft}}$
in the range 0.74--0.84~keV (A99),
and assumed solar abundances. Iron is the only element, whose
abundance in the warmer thermal component can be determined by our data.
We have fixed the relative abundances to the values 
reported in Buote et al. (1999) for all the elements to which the MECS is sensitive
(${\rm Z_{Si}}$:${\rm Z_S}$:${\rm Z_{Ar}}$:${\rm Z_{Ca}}$~=~1.16~:~1.04~:~0.72~:~0.99), and assumed solar values for the remainders.
None of the following results is substantially affected by variations
of the relative abundances within the statistical uncertainties in
Buote et al. (1999). 
The model is modified by Galactic photoelectric absorption
(${\rm N_H = 2.5 \times 10^{20}}$~cm$^{-2}$; Dickey \& Lockman 1990).
The results are summarized in Fig.~\ref{fig3} and Tab.~\ref{tab2}.
\begin{table}
\caption{Best-fit parameters and results for
the spatially-resolved MECS spectra (inner-outer radii of the extraction
annuli are indicated in the top row). All parameters derive
from the application of a two-temperature
plasma emission plus power-law model in the 2--10.5~keV band (details in text),
except ${\rm E_c}$, which is
the best-fit centroid of a Gaussian profile used to
fit the iron emission line complex. In the latter case, the fit is
performed in the 5--10.5~keV energy band, and the continuum is approximated
by a single temperature bremsstrahlung}
\begin{tabular*}{8.8cm}[h]{lcccc} \hline \hline
& 0$\arcmin$-2$\arcmin$ & 2$\arcmin$-4$\arcmin$ & 4$\arcmin$-6$\arcmin$ & 6$\arcmin$-8$\arcmin$ \\ \hline
kT$^a$  & $1.86 \pm^{0.40}_{0.11}$ & $2.31 \pm^{0.07}_{0.05}$ & $2.49 \pm 0.11$ & $2.9 \pm^{0.2}_{0.3}$ \\
${\rm Z_{Fe}}$ & $0.71 \pm^{0.19}_{0.18}$ & $0.57 \pm^{0.08}_{0.09}$ & $0.49 \pm^{0.09}_{0.19}$ & $0.29 \pm^{0.13}_{0.10}$ \\
${\rm \Delta \chi^2_{po}}$$^b$ & 17.2 & 0.3 & 0.0 & 0.0 \\
${\rm E_c}$$^a$ & $6.70 \pm^{0.05}_{0.04}$ & $6.67 \pm 0.03$ & $6.66 \pm^{0.05}_{0.04}$ & $6.70 \pm^{0.09}_{0.08}$ \\
$\chi^2/$dof & 43.4/33 & 37.2/33 & 36.7/33 & 43.8/29 \\ \hline \hline
\end{tabular*}

\noindent
$^a$in keV

\noindent
$^b$for the addition of the power-law to the two-temperature thermal model

\label{tab2}
\end{table}
The addition of a power-law model is required only for the innermost
spectrum. The ${\rm \Delta \chi^2}$ for the addition of two interesting
parameters is 17.2, significant at 98.5\% confidence level, according to
the F-test. ${\rm \Gamma}$ is only poorly constrained
(${\rm 0.9 \pm^{0.8}_{1.0}}$). If $\Gamma$ is fixed at 1.4 (1.7),
${\rm \Delta \chi^2 = 16.7}$ (15.6), significant at more than 99.9\%
level. No further spectral complexity
is required (if {\it e.g.} ${\rm kT_{soft}}$ is left free
${\rm \Delta \chi^2 < 0.1}$). If in the initial two thermal components model
both temperatures are left free to vary,
the r\^ole of the hard tail is played by
a hot plasma with ${\rm kT = 17 \pm^{\infty}_{13}}$~keV,
the temperature
of the component dominating the 2--7~keV flux is only marginally affected
(${\rm 1.77 \pm^{0.11}_{0.21}}$~keV),
and a slightly better $\chi^2$ is obtained (42.2/35~dof).
The addition of a hard power-law tail to the
other spectra yields no appreciable improvement in the quality of the
fits and the upper limits on its normalizations
are lower than the measure in the innermost spectrum at more than
90\% confidence level for two interesting parameters
(${\rm \Delta \chi^2 = 4.61}$; see Fig.~\ref{fig5}).
If a line corresponding to a fluorescent transition of neutral
iron were present, with Equivalent Width against the power-law of 200~eV
(as typically observed in Seyfert~1 galaxies, Nandra et al. 1997), we
would expect the measured centroid of the K$_{\alpha}$ line in the
innermost spectrum to be lower than in the others by about 130~eV, if the
neutral line is narrow,
and 90~eV, if broad ({\it i.e.}: intrinsic width $\sigma = 500$~eV).
Such variations are actually not observed.

\section{Discussion}

\begin{figure}
\includegraphics*[width=7.0cm,height=8.5cm,angle=-90]{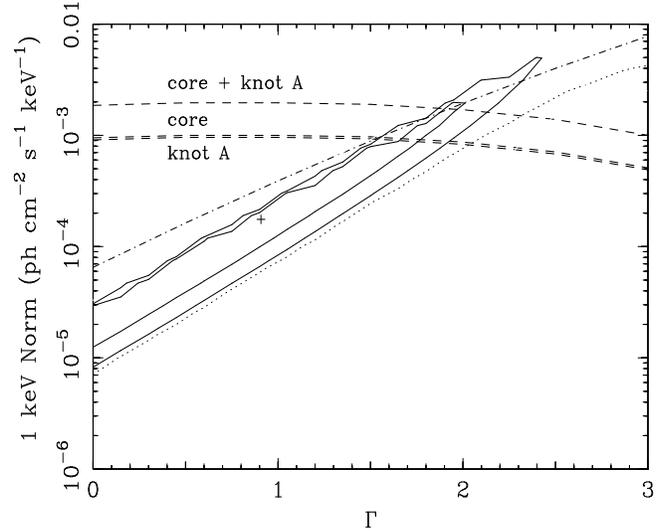}
\caption{{\it Solid lines}: iso-$\chi^2$ power-law normalization
versus photon index contours plot
(at 68\%, and 90\% for two interesting parameters) in the innermost 2$\arcmin$
around the M~87 core; {\it dotted line}: 90\%
normalization upper limit as a function of ${\rm \Gamma}$ for
the 2$\arcmin$--4$\arcmin$ spectrum. Similar curves for the
more external spectra are not shown for the sake of clarity; {\it dot-dashed
lines}: iso 2--10~keV flux curve for
$5 \times 10^{-12}$~erg~cm$^{-2}$~s$^{-2}$;
{\it dashed lines}:
locus of the points corresponding to the M~87 ROSAT/HRI quasi-simultaneous
observation fluxes. The core and ``knot A'' sources are shown separately
and summed}
\label{fig5}
\end{figure}

A hard X-ray power-law tail is statistically required only in the MECS
spectrum corresponding to the innermost 2$\arcmin$ from the M~87 core.
In Fig.~\ref{fig5} we show
the power-law normalization versus photon index iso-$\chi^2$
contours.
Although the strong correlation between these parameters
prevents us from setting any
meaningful constraints on them singularly,
the 2--10~keV flux is well bounded to be lower
than $6 \times 10^{-12}$~erg~s$^{-1}$~cm$^{-2}$ (at 90\%
confidence level for two interesting parameters). This is about a
factor 1.5 lower than that reported by A99,
confirming the variability
observed by Reynolds et al. (1999).
If this tail indeed originates in a low ${\rm \dot{m}}$
ADAF ($\Gamma = 1.4$) the corresponding bolometric X-ray luminosity
is $7.4 \times 10^{42}$~erg~s$^{-1}$. This implies ${\rm L/L_{Edd}
\approxlt 2 \times 10^{-5}}$, and ${\rm \dot{m} \sim 10^{-3}}$
assuming a viscosity parameter ${\rm \alpha = 0.3}$ (Esin et al. 1997;
Reynolds et al. 1997) and
a black hole mass of ${\rm 3 \times 10^9 M_{\odot}}$ (Macchetto et al. 1997).
A simple relation between the radio and X-ray output is observed
in low-luminosity active nuclei (Franceschini et al. 1998),
and can be explained from basic physical principles
(Yi \& Boughn 1999).
The ADAF 5~Ghz power predicted from our measure
(${\rm \sim 1.2 \times 10^{40} m^{8/5}_7 \dot{m}^{6/5}}$;
Yi \& Boughn 1998) is only a factor of a few lower than observed
(Franceschini et al. 1998),
a remarkable agreement given the simplicity of the model.

Other interpretations are, however, not ruled out by our data.
We have reanalyzed a public M~87 ROSAT/HRI
observation, which was performed about one week before the BeppoSAX
one. We followed
the reduction and analysis described in Harris et al. (1997) to
extract the fluxes from both the core and the knot A. In Fig~\ref{fig5}
we superpose
the locus of the spectral index versus normalization pairs corresponding
to the ROSAT/HRI fluxes and
the BeppoSAX contour plots for the same quantities,
assuming that the same power-law extends
from 0.1 to 10~keV and is modified only by the Galactic photoelectric
absorption (this plot is admittedly inspired to the Fig.~2 of
Reynolds et al. 1999).
If the hard X-ray power-law tail originates from one of the
two HRI sources, its spectral index must be comprised between
1.5 and 2.0. If the tail is instead given by the sum
of the two sources, the index of the superposed spectrum must
be comprised between 1.8 and 2.2.
These limits are not particularly demanding, being
consistent with the flat hard X-ray spectrum required in the
low ${\rm \dot{m}}$ ADAF
scenario, or with standard Seyfert-like spectra (Nandra et al.
1997). The lack of
any significant emissions from neutral iron fluorescent
transitions argues against the Seyfert scenario.
The central 2$\arcmin$ enclose much of the optical galaxy, the inner
radio lobes and the jet. It is rather unlikely that such a luminous
hard tail is produced by a collection of unresolved galactic sources,
but a substantial contribution from a jet cannot be ruled out. The
idea that M~87 may actually be a mis-aligned BLLac object was first
suggested by Tsvetanov et al. (1998) on the basis of HST data.
A complete characterization of the properties of the hard
tail in M~87 is an open challenge for
the forthcoming {\it Chandra} and XMM high resolution imaging capabilities.

\subsection{The thermal structure of the Virgo cluster core}

\begin{figure}
\includegraphics*[width=8.0cm,height=8.0cm]{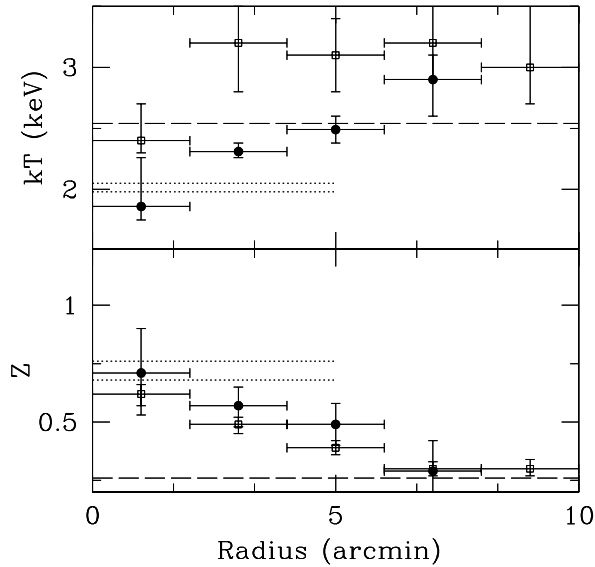}
\caption{Thermal plasma temperature ({\it upper panel}) and iron
abundance ({\it lower panel}) as a function of radius according to the
ASCA (Matsumoto et al. 1996; {\it empty squares})
and BeppoSAX (this {\it Letter}; {\it filled circles}) data.
The spatially-averaged values obtained from an analysis of the ASCA
innermost 5$\arcmin$ (A99; {\it dotted lines} bounded interval)
and RXTE (Reynolds et al. 1999; {\it dashed
line}) spectra are also shown}
\label{fig4}
\end{figure}

As a by-product of our analysis,
in Fig.~\ref{fig4} we compare the M~87 ``thermal structure''
obtained with
BeppoSAX, ASCA and RXTE. Preliminary results for the BeppoSAX data were
reported by D'Acri et al. (1998).
The BeppoSAX best-fit temperatures clearly increase with radius, whereas the
iron abundance decreases. While the latter result
reproduces the ASCA findings (Matsumoto et al. 1996), the former
represents a much clearer evidence than the original Matsumoto et al.
(1996) claim,
and is consistent with a gradual change along the distance from the
galaxy/cluster center. It is worth noting that the
ASCA Matsumoto et al. (1996) temperature measures are systematically
higher than the BeppoSAX ones. However, our spatially-averaged temperature
and iron abundance
are consistent with the
multi-phase analysis of both A99 (see Fig.~\ref{fig4})
and Buote et al. (1999). The relatively low abundance
measured by RXTE is likely to be due to the
contribution of the cluster emission
at larger radii in the 1$^{\circ}$ PCA field of view.

\begin{acknowledgements}
  
The BeppoSAX satellite is a joint Italian-Dutch program.
MG acknowledges an ESA Research Fellowship. The authors acknowledge
useful discussions with G.Matt, and gratefully thank an anonymous referee
for several comments, which strongly improved the quality of this {\it
Letter}

\end{acknowledgements}

\end{document}